\documentstyle[preprint,aps,epsf,floats]{revtex}

\tighten

\def\OMIT#1{{}}
\newcommand{\gsim}{\raisebox{-0.7ex}{$\stackrel{\textstyle >}{\sim}$ }}

\def\Dslash{D\hskip-0.65em /}
\def\Se{s}
\def\Tr{t}
\def\SD{x}
\begin{document}

\preprint{\vbox{
\hbox{NT@UW-01-016}
}}

\title{The Magnetic Moments of the Octet Baryons 
in Quenched Chiral Perturbation Theory}
\author{ Martin J. Savage\footnote{{\tt 
savage@phys.washington.edu}}}
\address{
Department of Physics, University of Washington, 
Seattle, WA 98195-1560.}

\maketitle

\begin{abstract} 
We compute the magnetic moments of the octet baryons up to two 
orders in quenched chiral perturbation theory.
In addition to the $\sim\sqrt{m_q}$ contributions that arise in QCD,
there are lower-order 
contributions of the form $M_0^2\log m_q$ from loop diagrams
involving hairpin interactions.
\end{abstract}

\bigskip
\vskip 8.0cm
\leftline{July 2001}

\vfill\eject

\section{Introduction}

The magnetic moments of the octet-baryons continue to play 
a crucial role in our understanding of hadronic structure
from the underlying theory of strong interactions, QCD, 
and also in the context of  more phenomenological models, 
such as the non-relativistic quark model (NRQM).
Perhaps one of the most intriguing aspects of strong interaction 
phenomenology is 
the remarkable success enjoyed by naive NRQM's
in describing the magnetic moments of the low-lying baryons 
based on the simple picture of weakly interacting 
Dirac particles with masses tuned to the baryon mass spectrum and 
one-third integer charges.
Clearly, one wishes to understand this success directly from QCD 
but at this point in time such illuminating calculations do not exist.

In the future, numerical lattice gauge calculations of 
many hadronic observables, including the magnetic moments, will be performed.
However, at present and in the near future, lattice calculations cannot be 
performed with the light quark masses,  $m_q$, near their physical values
($m_u\sim 5~{\rm MeV}$, $m_d\sim 10~{\rm MeV}$
and $m_s\sim 125~{\rm MeV}$)
and extrapolation from the lattice quark masses
($m_\pi^{\rm latt.}\sim 500~{\rm MeV}$)
is required.
First efforts at extrapolating 
quenched lattice evaluations of the octet magnetic
moments have been made in Refs.~\cite{MMThomas}.
Quenched lattice data~\cite{LWD91} was used in conjunction
with a Pade approximant to extrapolate
from $m_\pi^{\rm latt.}\gsim 600~{\rm MeV}$ to the physical value
of the pion mass.
Although the Pade approximant cannot be rigorously justified, 
it does reproduce  the correct small $m_q$ behavior~\cite{CP74}
($\sim \sqrt{m_q}$)
of the magnetic moments in QCD obtained from  chiral perturbation theory, 
and also the decoupling expected for large $m_\pi$, by construction.
This phenomenological extrapolation of the 
quenched data compares reasonably well
with the measured values of the magnetic moments
for many of the octet baryons, however, there are still
significant discrepancies that remain.
One avenue that could be explored is to include the higher order $m_q$
dependences that have been computed during the past several 
years~\cite{JLMS92,MS97,DH98,PRM00} into the Pade approximant.
However, given that quenched QCD (QQCD) is not QCD,
and that the Pade approximant cannot be justified from QCD
it is perhaps more interesting to know the predictions of QQCD itself
for the magnetic moments, and not to attempt to match QQCD to the real world.
Such predictions would require calculations in QQCD with quark masses that
are small enough to match onto the chiral expansion
and also a calculation of the magnetic moments in quenched chiral perturbation
theory (Q$\chi$PT)~\cite{Sharpe90,S92,BG92,LS96}, 
which is what we will do here.

In QCD the expansion 
of a baryon magnetic moment $\mu_B$ 
about the chiral limit has the form
$\mu_B\ \sim\ \mu_0\ +\ \beta\  \sqrt{m_q}\ +\ \gamma\  m_q\log m_q\ +\ ...$,
where $\mu_0$ is a combination of the two 
coefficients of  local dimension-5 operators that contribute at leading order
in the chiral expansion.
However, in QQCD the presence of the light $\eta^\prime$, and the associated
``hairpin'' interactions render QQCD more divergent in the infrared than
QCD.
Therefore, in QQCD the chiral expansion for the magnetic
moments of the low-lying baryons is of the form
$\mu_B^{\rm QQCD}\ \sim\ \mu_0^{\rm QQCD}\ +\ \delta\ \log m_q
\ +\ \beta^{\rm QQCD}\  \sqrt{m_q}\ +\ ...$,
where $ \mu_0^{\rm QQCD} \ne \mu_0$, and 
$\beta^{\rm QQCD}\ne\beta$. 
Further, the coefficient $\delta$ is entirely an artifact of quenching.
In this work we compute the magnetic moments of the octet-baryons up to two
orders in Q$\chi$PT, that is we compute $\delta$ and $\beta^{\rm QQCD}$
for each baryon.

Q$\chi$PT for the low-lying baryons has been set-up 
in a pioneering paper by 
Labrenz and Sharpe (LS)~\cite{LS96} and used to compute the chiral 
corrections to the octet and decuplet baryon masses.
LS used quark-line diagrams to identify and compute the meson-loop diagrams
that contribute.
In this work, we 
will use ghost-baryon fields 
to compute the one-loop diagrams that contribute to the magnetic moments, 
in a way similar to that used to compute properties of baryons 
containing heavy  quarks~\cite{Chil97}.

\section{Q$\chi$PT}

The lagrange density of QQCD is
\begin{eqnarray}
{\cal L} & = & 
\sum_{a,b=u,d,s} \overline{q}^a\ 
\left[\ i\Dslash -m_{q}\ \right]_a^b\ q_b
\ +\ 
\sum_{\tilde a,\tilde b=\tilde u, \tilde d,\tilde s}
\overline{\tilde q}^{\tilde a}
\ \left[\ i\Dslash -m_{\tilde q}\ 
\right]_{\tilde a}^{\tilde b}\tilde q_{\tilde b}
\nonumber\\
& = & 
\sum_{j,k=u,d,s,\tilde u, \tilde d,\tilde s} 
\overline{Q}^j\ 
\left[\ i\Dslash -m_{Q}\ \right]_j^k\ Q_k
\ \ \ ,
\label{eq:QQCD}
\end{eqnarray}
where $q_a$ are the three light-quarks, $u$, $d$, and $s$, and 
$\tilde q$ are three light bosonic quarks  $\tilde u$, $\tilde d$, 
and $\tilde s$.
The super-quark field, $Q_j$, is a six-component column vector with the 
three light-quarks, $u$, $d$, and $s$,
in the upper three entries and the three ghost-light-quarks, 
$\tilde u$, $\tilde d$, and $\tilde s$,
in the lower three entries.
The graded equal-time commutation relations for two fields is
\begin{eqnarray}
Q_i^\alpha ({\bf x}) Q_j^{\beta \dagger} ({\bf y}) - 
(-)^{\eta_i\eta_j}Q_j^{\beta \dagger} ({\bf y})Q_i^\alpha ({\bf x})
& = & 
\delta^{\alpha\beta}\delta_{ij}\delta^3({\bf x}-{\bf y})
\ \ \ ,
\label{eq:comm}
\end{eqnarray}
where $\alpha,\beta$ are Dirac-indices and $i,j$ are flavor indices.
The objects $\eta_k$ correspond to the parity of the component of $Q_k$,
with $\eta_k=+1$ for $k=1,2,3$ and $\eta_k=0$ for $k=4,5,6$, and 
the graded commutation relations for two $Q$'s or two $Q^\dagger$'s
are analogous.
The diagonal super mass-matrix, $m_Q$, has entries
$m_Q = {\rm diag}(m_u,m_d,m_s,m_u,m_d,m_s)$,
i.e.  
$m_{\tilde u}=m_u$, $m_{\tilde d}=m_d$ and 
$m_{\tilde s}=m_s$, so that the contribution 
to the determinant in the path integral 
from the $q$'s and the $\tilde q$'s exactly cancel.

In the absence of quark masses,
the lagrange density in eq.~(\ref{eq:QQCD})
has a
graded symmetry $U(3|3)_L\otimes U(3|3)_R$, where the left- and 
right-handed quark fields transform as
$Q_L\rightarrow U_L Q_L$ and $Q_R\rightarrow U_R Q_R$ respectively.
However, the functional integral associated with this Lagrange density 
does not converge unless the transformations on the left- and right-handed
fields are related, ${\rm sdet}(U_L)={\rm sdet}(U_R)$, 
where ${\rm sdet}()$ denotes a superdeterminant~\cite{S92,BG92,DOTV98,SS01},
leaving the theory to have a symmetry
$\left[SU(3|3)_L\otimes SU(3|3)_R\right]\times U(1)_V$, 
where the ``$\times$'' denotes  a semi-direct product as opposed 
to a direct product, ``$\otimes$''~\footnote{
A simple example of the implications of this distinction can be
seen by considering the graded group $U(1|1)$~\cite{Guhr93}.}.
It is assumed that this symmetry is spontaneously broken 
$\left[SU(3|3)_L\otimes SU(3|3)_R\right]\times U(1)_V\rightarrow 
SU(3|3)_V\times U(1)_V$ so that an identification with QCD can be made.

\subsection{The Pseudo-Goldstone Bosons}

The pseudo-Goldstone bosons of QQCD form a $6\times 6$ matrix, $\Phi$, 
that can be written in block form
\begin{eqnarray}
\Phi & = & \left(\matrix{\pi & \chi^\dagger \cr \chi & \tilde\pi}\right)
\ \ \ ,
\label{eq:mesondef}
\end{eqnarray}
where $\pi$ is the $3\times 3$ matrix of pseudo-Goldstone bosons 
including the $\eta^\prime$
with quantum numbers of $\overline{q}q$ pairs,
$\tilde \pi$ is a  $3\times 3$ matrix of pseudo-Goldstone bosons 
including the $\tilde\eta^\prime$
with quantum numbers of $\overline{\tilde q}\tilde q$ pairs, and 
$\chi$ is a $3\times 3$ matrix of pseudo-Goldstone fermions
with quantum numbers of $\overline{\tilde q}q$ pairs,
\begin{eqnarray}
\pi & = & \left(\matrix{\eta_u &\pi^+ &K^+ \cr \pi^- &\eta_d & K^0\cr
K^- &\overline{K}^0 & \eta_s}\right)
\ \ ,\ \ 
\tilde\pi \ = \ \left(\matrix{\tilde\eta_u &\tilde\pi^+ &
\tilde K^+ \cr \tilde\pi^- &\tilde\eta_d & \tilde K^0\cr
\tilde K^- &\overline{\tilde K}^0 & \tilde\eta_s}\right)
\ \ ,\ \ 
\chi\ =\  \left(\matrix{\chi_{\eta_u} &\chi_{\pi^+} &\chi_{K^+} \cr 
\chi_{\pi^-} &\chi_{\eta_d} & \chi_{K^0}\cr
\chi_{K^-} &\chi_{\overline{K}^0} &\chi_{\eta_s}}\right)
\ \ ,
\label{eq:mesmats}
\end{eqnarray}
As the object 
\begin{eqnarray}
\Phi_0 & = & {1\over\sqrt{6}}{\rm str}\left(\Phi\right)
\ =\ {1\over\sqrt{2}}\left(\ \eta^\prime - \tilde\eta^\prime\ \right)
\ \ ,
\label{eq:phi0}
\end{eqnarray}
is invariant under 
$\left[SU(3|3)_L\otimes SU(3|3)_R\right]\times U(1)_V$
(but not under $U(3|3)_L\otimes U(3|3)_R$) 
the most general lagrange density that describes low-momentum 
dynamics will contain arbitrary functions of 
$\Phi_0$~\cite{S92,BG92}.
At lowest order in the chiral expansion, the Lagrange density
that describes the dynamics of the pseudo-Goldstone bosons 
is, using the notation of Ref.~\cite{LS96},
\begin{eqnarray}
{\cal L} & = & 
{f^2\over 8}\ {\rm str}\left[\  
\partial^\mu\Sigma^\dagger\partial_\mu\Sigma\ \right]
\ +\ 
\lambda\ {\rm str} \left[\ m_Q\Sigma\ +\ m_Q^\dagger\Sigma^\dagger\ \right]
\ +\ \alpha_\Phi \ \partial^\mu\Phi_0\partial_\mu\Phi_0
\ -\ 
M_0^2\  \Phi_0^2
\ \ \ ,
\label{eq:lagpi}
\end{eqnarray}
where the parameter $\lambda$ is chosen to reproduce the meson masses,
and $\Sigma$ is the exponential of the $\Phi$ field,
\begin{eqnarray}
\Sigma & = & {\rm exp}\left( {2\  i\  \Phi\over f}\right)
\ \ .
\label{eq:sigdef}
\end{eqnarray}
With this normalization, $f\sim 132~{\rm MeV}$ in QCD.
In addition, 
it is understood that the operators with coefficients $\alpha_\Phi$ and
$M_0$, the hairpin interactions,
are inserted perturbatively.
Expanding out the Lagrange density in eq.~(\ref{eq:lagpi}) to quadratic order
in the meson fields, one finds relations between the meson masses in the
isospin
limit,
\begin{eqnarray}
m_{\eta_s}^2 & = & 2 m_K^2 - m_\pi^2
\ \ ,\ \ 
m_{\eta_u}^2 \ = \ m_{\eta_d}^2\ =\ m_\pi^2
\ \ \ .
\label{eq:massrels}
\end{eqnarray}
The Lagrange density in eq.~(\ref{eq:lagpi})
has been used to compute several observables in the
meson sector, such as $f_K$, $f_\pi$~\cite{S91} 
and the meson masses out to one-loop in
perturbation theory~\cite{S92,BG92,Pall97} 
(for a nice review see Ref.~\cite{G94}).

\subsection{The Octet Baryons}

The inclusion of the lowest-lying baryons, 
the octet of spin-${1\over 2}$ baryons 
and the decuplet of spin-${3\over 2}$ baryon resonances,
is detailed in Ref.~\cite{LS96}.
An interpolating field that has non-zero overlap with the 
baryon octet (when the $ijk$ indices are restricted to $1,2,3$) 
is~\cite{LS96}
\begin{eqnarray}
{\cal B}^\gamma_{ijk} & \sim &
\left[\ Q_i^{\alpha,a} Q_j^{\beta,b} Q_k^{\gamma,c}
\ -\  Q_i^{\alpha,a} Q_j^{\gamma,c} Q_k^{\beta,b}\ \right]
\epsilon_{abc} \left(C\gamma_5\right)_{\alpha\beta}
\ \ \ ,
\label{eq:octinter}
\end{eqnarray}
where $C$ is the charge conjugation operator,
$a,b,c$ are color indices and $\alpha,\beta,\gamma$ are Dirac-indices.
Dropping the Dirac-index, one finds that
under the interchange of flavor indices~\cite{LS96},
\begin{eqnarray}
{\cal B}_{ijk} & = & (-)^{1+\eta_j \eta_k}\  {\cal B}_{ikj}
\ \ ,\ \ 
{\cal B}_{ijk} \ +\  (-)^{1+\eta_i \eta_j}\ {\cal B}_{jik}
\ +\ (-)^{1 + \eta_i\eta_j + \eta_j\eta_k + \eta_k\eta_i}\ 
{\cal B}_{kji}\ =\ 0
\ \ \ .
\label{eq:bianchi}
\end{eqnarray}
In analogy with QCD, 
we consider the transformation of $B_{ijk}$ under $SU(3|3)_V$
transformations, and using the graded relation
\begin{eqnarray}
Q_i\ U^j_{\ k} & = & (-)^{\eta_i (\eta_j+\eta_k)} \ U^j_{\ k}\  Q_i
\ \ ,
\label{eq:gradtrans}
\end{eqnarray}
in eq.~(\ref{eq:octinter}),
it is straightforward to show that~\cite{LS96}
\begin{eqnarray}
{\cal B}_{ijk} & \rightarrow & 
(-)^{\eta_l (\eta_j+\eta_m) +(\eta_l+\eta_m)(\eta_k+\eta_n)} 
\ U_i^{\ l}\  U_j^{\ m}\  U_k^{\ n}\ 
{\cal B}_{lmn}
\ \ \ .
\label{eq:octtrans}
\end{eqnarray}

The object ${\cal B}_{ijk}$ describes a {\bf 70} dimensional representation
of $SU(3|3)_V$~\cite{LS96}.
It is convenient to 
decompose the irreducible representations of $SU(3|3)_V$ 
into  irreducible representations
$SU(3)_{q}\otimes SU(3)_{\tilde q}\otimes U(1)$~\cite{BBI81,BB81,HM83}, 
and we will forget about the $U(1)$'s from now on.
The subscript denotes where the $SU(3)$ acts, either on the $q$'s or on the
$\tilde q$'s.
The ground floor of the {\bf 70}-dimensional representation contains baryons
that are comprised of three quarks, $qqq$, and is therefore an 
$({\bf 8},{\bf 1})$ of  $SU(3)_{q}\otimes SU(3)_{\tilde q}$.
It was shown in LS that the octet baryons are embedded as
\begin{eqnarray}
{\cal B}_{abc} & = & {1\over\sqrt{6}}
\left( \ \epsilon_{abd}\ B^d_c\ +\ 
\epsilon_{acd} B^d_b\ \right)
\ \ \ ,
\end{eqnarray}
where the indices are restricted to take the values $a,b,c=1,2,3$ only.
The octet baryon matrix is
\begin{eqnarray}
B & = & \left(\matrix{{1\over\sqrt{6}}\Lambda + {1\over\sqrt{2}}\Sigma^0
& \Sigma^+ & p\cr
\Sigma^- & {1\over\sqrt{6}}\Lambda - {1\over\sqrt{2}}\Sigma^0 & n\cr
\Xi^- & \Xi^0 & -{2\over\sqrt{6}}\Lambda}\right)
\ \ \ .
\label{eq:baryons}
\end{eqnarray}
The first floor of the  {\bf 70}-dimensional representation contains baryons
that are composed of two quarks and one ghost-quark, $\tilde q qq$,
and therefore transforms as 
$({\bf 6},{\bf 3})\oplus (\overline{\bf 3},{\bf 3})$
of  $SU(3)_{q}\otimes SU(3)_{\tilde q}$.
The tensor representation $_{\tilde a} \Se_{ab}$ 
of the $({\bf 6},{\bf 3})$ multiplet, where ${\tilde a}=1,2,3$ runs over the 
$\tilde q$ indices and $a,b=1,2,3$ run over the $q$ indices,
has baryon assignment
\begin{eqnarray}
_{\tilde a} \Se_{11} & = & 
\Sigma_{\tilde a}^{+1}
\ \ ,\ \ 
_{\tilde a} \Se_{12}\ =\ _{\tilde a} \Se_{21}\ =\ 
{1\over\sqrt{2}} \Sigma_{\tilde a}^{0}
\ \ ,\ \ 
_{\tilde a} \Se_{22}\ =\ 
\Sigma_{\tilde a}^{-1}
\nonumber\\
_{\tilde a} \Se_{13}& = & _{\tilde a} \Se_{31}
\ = \
{1\over\sqrt{2}}
\ ^{(6)}\Xi_{\tilde a}^{+{1\over 2}}
\ \ ,\ \ 
_{\tilde a} \Se_{23}\ =\ _{\tilde a} \Se_{32}
\ =\ 
{1\over\sqrt{2}}
\ ^{(6)}\Xi_{\tilde a}^{-{1\over 2}}
\ \ ,\ \ 
_{\tilde a} \Se_{33} \ =\ 
\Omega_{\tilde a}^0
\ \ \ ,
\label{eq:sixdef}
\end{eqnarray}
where the notation closely follows that used to describe baryons containing a 
single heavy quark~\footnote{Heavy baryons are classified  
by irreducible representations of $SU(2)_H\otimes SU(3)_V$, where $SU(2)_H$ is
the $c,b$ heavy-quark-flavor symmetry group and 
$SU(3)_V$ is the light-quark flavor symmetry group.}.
The right superscript denotes the third component of q-isospin,
while the left subscript denotes the $\tilde q$ flavor.
The tensor representation $_{\tilde a} \Tr^a$ 
of the $(\overline{\bf 3},{\bf 3})$ 
multiplet, where ${\tilde a}=1,2,3$ runs over the 
$\tilde q$ indices and $a=1,2,3$ run over the $q$ indices,
has baryon assignment
\begin{eqnarray}
_{\tilde a} \Tr^1 & = & ^{(\overline{3})}\Xi_{\tilde a}^{-{1\over 2}}
\ \ ,\ \ 
_{\tilde a} \Tr^2 \ =\  ^{(\overline{3})}\Xi_{\tilde a}^{+{1\over 2}}
\ \ ,\ \ 
_{\tilde a} \Tr^3 \ =\  \Lambda_{\tilde a}^0
\ \ \ .
\label{eq:tripdef}
\end{eqnarray}
The $_{\tilde a} \Se_{ab}$ and the $_{\tilde a} \Tr^a$ 
are uniquely embedded into
${\cal B}_{ijk}$ (up to field redefinitions), 
constrained by the relations in eq.~(\ref{eq:bianchi}):
\begin{eqnarray}
{\cal B}_{ijk} & = & 
\sqrt{2\over 3\ }\ _{i-3}\Se_{jk}
\ \ \ \ \ {\rm for}\  \ \ \ i=4,5,6\ \ {\rm and}\ \ j,k=1,2,3
\nonumber\\
{\cal B}_{ijk} & = & 
{1\over 2}\ \  _{j-3}\Tr^\sigma \varepsilon_{\sigma i k }
\ +\ {1\over\sqrt{6}}\ \ _{j-3}\Se_{ik}
\ \ \ \ \ {\rm for}\  \ \ \ j=4,5,6\ \ {\rm and}\ \ i,k,\sigma =1,2,3
\nonumber\\
{\cal B}_{ijk} & = & 
-{1\over 2}\ \   _{k-3}\Tr^\sigma \varepsilon_{\sigma i j }
\ -\ {1\over\sqrt{6}}\ \  _{k-3}\Se_{ij}
\ \ \ \ \ {\rm for}\  \ \ \ k=4,5,6\ \ {\rm and}\ \ i,j,\sigma =1,2,3
\ \ \ .
\label{eq:firstfloor}
\end{eqnarray}
As we are only interested in one-loop contributions to observables with 
$qqq$-baryons in the asymptotic states, we do not 
explicitly construct the second and 
third floors of the {\bf 70}.
However, it is straightforward to show that the 
second floor, consisting of $\tilde q\tilde q q$-baryons,
can be decomposed into
$({\bf 3},{\bf 6})\oplus ({\bf 3},\overline{\bf 3})$
of  $SU(3)_{q}\otimes SU(3)_{\tilde q}$, while the third floor,
consisting of $\tilde q\tilde q\tilde q$-baryons can be decomposed into
a $({\bf 1},{\bf 8})$ of  $SU(3)_{q}\otimes SU(3)_{\tilde q}$.
To make contact with the discussions by LS~\cite{LS96}, 
one can consider how the four
floors transform under the vector $q+\tilde q$ symmetry of
$SU(3)_q\otimes SU(3)_{\tilde q} \equiv
SU(3)_{q+\tilde q}\otimes SU(3)_{q-\tilde q}$.
The ground floor transforms as an ${\bf 8}$ of $SU(3)_{q+\tilde q}$,
the first floor transforms as an 
$({\bf 3}\otimes {\bf 6})\oplus ( {\bf 3}\otimes\overline{\bf 3}) =
{\bf 10}\oplus{\bf 8}\oplus{\bf 8}\oplus{\bf 1}$, similarly
the second floor transforms as an 
${\bf 10}\oplus{\bf 8}\oplus{\bf 8}\oplus{\bf 1}$,
and the third floor transforms as an ${\bf 8}$.

\subsection{The Decuplet Baryons}

As the mass splitting between the 
decuplet and octet baryons
(in QCD) is much less than the scale of chiral
symmetry breaking ($\Lambda_\chi\sim 1~{\rm GeV}$)
the decuplet must be included as a dynamical field in order to have a
theory where the natural scale of higher order interactions is set by
$\Lambda_\chi$.
We assume that the decuplet-octet mass splitting is still small 
compared to the 
scale of chiral symmetry breaking in QQCD.
An interpolating field that contains the spin-${3\over 2}$
decuplet as the ground floor is~\cite{LS96}
\begin{eqnarray}
{\cal T}^{\alpha ,\mu}_{ijk} & \sim &
\left[
Q^{\alpha,a}_i Q^{\beta,b}_j Q^{\gamma,c}_k +
Q^{\beta,b}_i Q^{\gamma,c}_j Q^{\alpha,a}_k  +
Q^{\gamma,c}_i Q^{\alpha,a}_j Q^{\beta,b}_k 
\right]
\varepsilon_{abc} (C\gamma^\mu)_{\beta\gamma}
\ \ \ ,
\label{eq:tdef}
\end{eqnarray}
where the indices $i,j,k$ run from $1$ to $6$.
Neglecting Dirac-indices, one finds that
under the interchange of flavor indices~\cite{LS96}
\begin{eqnarray}
{\cal T}_{ijk} & = & 
(-)^{1+\eta_i\eta_j} {\cal T}_{jik}\ =\ 
(-)^{1+\eta_j\eta_k} {\cal T}_{ikj}
\ \ \ .
\label{eq:ttrans}
\end{eqnarray}

${\cal T}_{ijk}$ describes a ${\bf 38}$ dimensional representation
of $SU(3|3)_V$, which has a ground floor transforming as 
$({\bf 10},{\bf 1})$ under $SU(3)_q\otimes SU(3)_{\tilde q}$ with
\begin{eqnarray}
{\cal T}_{abc} & = & T_{abc}
\ \ \ ,
\label{eq:Tbarys}
\end{eqnarray}
where the indices are restricted to take the values $a,b,c=1,2,3$,
and where $T_{abc}$ is the totally symmetric tensor containing
the decuplet of baryon resonances,
\begin{eqnarray}
T_{111} & = & \Delta^{++}
\ \ ,\ \ 
T_{112} \ =\  {1\over\sqrt{3}}\Delta^+
\ \ ,\ \ 
T_{122} \ =\  {1\over\sqrt{3}}\Delta^0
\ \ ,\ \ 
T_{222} \ =\   \Delta^{-}
\nonumber\\
T_{113} & = & {1\over\sqrt{3}}\Sigma^{*,+}
\ \ ,\ \ 
T_{123} \ =\  {1\over\sqrt{6}}\Sigma^{*,0}
\ \ ,\ \ 
T_{223} \ =\  {1\over\sqrt{3}}\Sigma^{*,-}
\nonumber\\
T_{133} & = & {1\over\sqrt{3}}\Xi^{*,0}
\ \ ,\ \ 
T_{233} \ =\  {1\over\sqrt{3}}\Xi^{*,-}
\ \ ,\ \ 
T_{333} \ =\  \Omega^{-}
\ \ \ .
\label{eq:decuplet}
\end{eqnarray}
The first floor transforms as a $({\bf 6},{\bf 3})$ under
$SU(3)_q\otimes SU(3)_{\tilde q}$ which has a tensor representation,
$_{\tilde a} \SD_{ij}$, with baryon assignment
\begin{eqnarray}
_{\tilde a} \SD_{11} & = & 
\Sigma_{\tilde a}^{*,+1}
\ \ ,\ \ 
_{\tilde a} \SD_{12}\ =\ _{\tilde a} \SD_{21}\ =\ 
{1\over\sqrt{2}} \Sigma_{\tilde a}^{*,0}
\ \ ,\ \ 
_{\tilde a} \SD_{22}\ =\ 
\Sigma_{\tilde a}^{*,-1}
\nonumber\\
_{\tilde a} \SD_{13} & = & _{\tilde a} \SD_{31}
\ =\ 
{1\over\sqrt{2}}
\ \Xi_{\tilde a}^{*,+{1\over 2}}
\ \ ,\ \ 
_{\tilde a} \SD_{23}\ =\ _{\tilde a} \SD_{32}
\ =\ 
{1\over\sqrt{2}}
\ \Xi_{\tilde a}^{*,-{1\over 2}}
\ \ ,\ \ 
_{\tilde a} \SD_{33} \ =\  
\Omega_{\tilde a}^{*,0}
\ \ \ .
\label{eq:sixTdef}
\end{eqnarray}
The embedding of $_{\tilde a} \SD_{ij}$ into ${\cal T}_{ijk}$ is 
unique (up to field redefinitions),
constrained by the symmetry properties in eq.~(\ref{eq:ttrans}):
\begin{eqnarray}
{\cal T}_{ijk} & = & 
+ {1\over\sqrt{3}}\ _{i-3}\SD_{jk}
\ \ \ \ \ {\rm for}\  \ \ \ i=4,5,6\ \ {\rm and}\ \ j,k=1,2,3
\nonumber\\
{\cal T}_{ijk} & = & 
-{1\over\sqrt{3}}\ \ _{j-3}\SD_{ik}
\ \ \ \ \ {\rm for}\  \ \ \ j=4,5,6\ \ {\rm and}\ \ i,k =1,2,3
\nonumber\\
{\cal T}_{ijk} & = & 
+ {1\over\sqrt{3}}\ \  _{k-3}\SD_{ij}
\ \ \ \ \ {\rm for}\  \ \ \ k=4,5,6\ \ {\rm and}\ \ i,j =1,2,3
\ \ \ .
\label{eq:firstfloorT}
\end{eqnarray}

Again, we do not explicitly
construct the second and third floor baryons of the {\bf 38}.
However, it is easy to show that the second floor transforms as a 
$({\bf 3},\overline{\bf 3})$ and that the third floor transforms as a 
$({\bf 1},{\bf 1})$ under $SU(3)_q\otimes SU(3)_{\tilde q}$.
Considering the transformation of the floors under 
$SU(3)_{q+\tilde q}$ we recover the results of LS, where the ground floor
transforms as a ${\bf 10}$, the first floor transforms as a ${\bf 10}\oplus
{\bf 8}$,
the second floor transforms as a ${\bf 8}\oplus {\bf 1}$ and the third floor
transforms as a ${\bf 1}$.

\subsection{Lagrange Density for the Baryons}

The free Lagrange density for the ${\cal B}_{ijk}$ and 
${\cal T}_{ijk}$ fields is, at leading order in the heavy baryon 
expansion~\cite{JMheavy,JMaxial,Jmass,chiralN,chiralUlf}, 
and using the notation of LS~\cite{LS96},
\begin{eqnarray}
{\cal L} & = & 
i\left(\overline{\cal B} v\cdot {\cal D} {\cal B}\right)
\ +\ 2\alpha_M \left(\overline{\cal B}{\cal B}{\cal M}_+\right)
\ +\ 2\beta_M \left(\overline{\cal B}{\cal M}_+{\cal B}\right)
\ +\ 2\sigma_M \left(\overline{\cal B}{\cal B}\right)\ 
{\rm str}\left({\cal M}_+\right)
\nonumber\\
& - & 
i \left(\overline{\cal T}^\mu v\cdot {\cal D} {\cal T}_\mu\right)
\ +\ 
\Delta\ \left(\overline{\cal T}^\mu {\cal T}_\mu\right)
\ +\ 2\gamma_M \left(\overline{\cal T}^\mu{\cal M}_+{\cal T}_\mu\right)
\ -\ 2 \overline{\sigma}_M  \left(\overline{\cal T}^\mu {\cal T}_\mu\right)\
{\rm str}\left({\cal M}_+\right)
\ \ ,
\label{eq:free}
\end{eqnarray}
where 
${\cal M}_+={1\over 2}\left(\xi^\dagger m_Q\xi^\dagger + \xi m_Q\xi\right)$
(which differs by a factor of $2$ from LS),
$\Delta$ is the decuplet-octet mass splitting,
and $\xi=\sqrt{\Sigma}$.
The brackets, $\left(\ \right)$ denote contraction of lorentz and flavor
indices as defined in LS~\cite{LS96}.
For a matrix $\Gamma^\alpha_\beta$ acting in spin-space, 
and a matrix $Y_{ij}$
acting in flavor-space (with or without lorentz indices), 
the required contractions are~\cite{LS96}
\begin{eqnarray}
\left(\overline{\cal B}\  \Gamma \ {\cal B}\right)
& = & 
\overline{\cal B}^{\alpha,kji}\ \Gamma_\alpha^\beta\  {\cal B}_{ijk,\beta}
\ \ ,\ \ 
\left(\overline{\cal T}^\mu\  \Gamma \ {\cal T}_\mu\right)
\ =\ 
\overline{\cal T}^{\mu\alpha,kji}\ \Gamma_\alpha^\beta\  
{\cal T}_{ijk,\beta\mu}
\nonumber\\
\left(\overline{\cal B}\  \Gamma \ Y\ {\cal B}\right)
& = & 
\overline{\cal B}^{\alpha,kji}\ \Gamma_\alpha^\beta\  
Y_i^{\ l}\ 
{\cal B}_{ljk,\beta}
\ \ ,\ \ 
\left(\overline{\cal T}^\mu\  \Gamma \ Y\ {\cal T}_\mu\right)
\ =\ 
\overline{\cal T}^{\mu\alpha,kji}\ \Gamma_\alpha^\beta\  
Y_i^{\ l}\ 
{\cal T}_{ljk,\beta\mu}
\nonumber\\
\left(\overline{\cal B}\  \Gamma \ {\cal B}\ Y\right)
& = & 
(-)^{(\eta_i+\eta_j)(\eta_k+\eta_n)}
\overline{\cal B}^{\alpha,kji}\ \Gamma_\alpha^\beta\  
Y_k^{\ n}\ 
{\cal B}_{ijn,\beta}
\nonumber\\
\left(\overline{\cal B}\  \Gamma \ Y^\mu {\cal T}_\mu\right)
& = & 
\overline{\cal B}^{\alpha,kji}\ \Gamma_\alpha^\beta\  
\left(Y^\mu\right)_i^l
{\cal T}_{ljk,\beta\mu}
\ \ \ ,
\label{eq:Contractions}
\end{eqnarray}
where $\overline{\cal B}$ and $\overline{\cal T}$
transform the same way,
\begin{eqnarray}
\overline{B}^{kji}&\rightarrow &
(-)^{\eta_l (\eta_j+\eta_m) +(\eta_l+\eta_m)(\eta_k+\eta_n)} 
\overline{\cal B}^{nml}
 U_{n}^{\ k\dagger}\ U_{m}^{\ j\dagger}\ U_{l}^{\ i\dagger} \  
\ \ \ .
\end{eqnarray}

The Lagrange density describing the interactions of the baryons with the
pseudo-Goldstone bosons is~\cite{LS96}
\begin{eqnarray}
{\cal L} & = & 
2\alpha\ \left(\overline{\cal B} S^\mu {\cal B} A_\mu\right)
\ +\ 
2\beta\ \left(\overline{\cal B} S^\mu A_\mu {\cal B} \right)
\ +\ 
2\gamma\ \left(\overline{\cal B} S^\mu {\cal B} \right)
\ {\rm str}\left(A_\mu \right)
\nonumber\\
+ & &  
2{\cal H} \left(\overline{\cal T}^\nu S^\mu A_\mu {\cal T}_\nu \right)
\ +\ 
\sqrt{3\over 2}{\cal C} 
\left[\ 
\left( \overline{\cal T}^\nu A_\nu {\cal B}\right)\ +\ 
\left({\cal B} A_\nu {\cal T}^\nu\right)\ \right]
\ +\ 2\gamma^\prime
\left(\overline{\cal T}^\nu S^\mu {\cal T}_\nu\right) 
\ {\rm str}\left(A_\mu \right)
\ ,
\label{eq:ints}
\end{eqnarray}
where $S^\mu$ is the covariant spin-vector~\cite{JMheavy,JMaxial,Jmass}, and 
where the sign of the ${\cal C}$-term is opposite to that of LS~\footnote{
The sign of ${\cal C}$ has no physical meaning as  a field redefinition
on either baryon field, ${\cal B}\rightarrow -{\cal B}$ or 
${\cal T}\rightarrow -{\cal T}$ changes the sign of the ${\cal C}$-term.
}.
Restricting oneself to the $qqq$ sector, it is straightforward to show
that 
\begin{eqnarray}
\alpha & =& {2\over 3}D+2F
\ \ ,\ \ 
\beta \ =\ -{5\over 3}D + F 
\ \ ,
\label{eq:couplings}
\end{eqnarray}
where $D$ and $F$ are constants that multiply the $SU(3)_q$ invariants
that are commonly used in
QCD, and it should be stressed that the $F$ and $D$ discussed here 
will not have the numerical values of those of QCD.
In QQCD there is an additional coupling, $\gamma$,
a hairpin interaction~\cite{LS96} that is not usually considered
in the $\chi$PT description of low-energy QCD.
In our calculation of the magnetic moments we will
replace $\alpha$ and $\beta$ with $F$ and $D$, but we will keep $\gamma$ 
explicit, as in LS~\cite{LS96}.
In the above discussion, vector and axial-vector meson
fields have been introduced in direct analogy with QCD.
The covariant derivative acting on either the ${\cal B}$ or ${\cal T}$ fields
has the form
\begin{eqnarray}
\left({\cal D}^\mu{\cal B}\right)_{ijk} & = & 
\partial^\mu {\cal B}_{ijk}
+
\left(V^\mu\right)^l_i {\cal B}_{ljk}
+ 
(-)^{\eta_i (\eta_j+\eta_m)} \left(V^\mu\right)^m_j {\cal B}_{imk}
+ (-)^{(\eta_i+\eta_j) (\eta_k+\eta_n)}
\left(V^\mu\right)^n_k {\cal B}_{ijn}
\ ,
\label{eq:covariant}
\end{eqnarray}
where the vector and axial-vector meson fields are
\begin{eqnarray}
V^\mu & = & {1\over 2}\left(\ \xi\partial^\mu\xi^\dagger
\ + \ 
\xi^\dagger\partial^\mu\xi \ \right)
\ \ ,\ \ 
A^\mu \ =\  {i\over 2}\left(\ \xi\partial^\mu\xi^\dagger
\ - \ 
\xi^\dagger\partial^\mu\xi \ \right)
\ \ \ .
\label{eq:mesonfields}
\end{eqnarray}

It is useful to expand the interactions of the ground and first
floors,
\begin{eqnarray}
\left( \overline{\cal B} S^\mu {\cal B} A_\mu \right) & = & 
{1\over f}
\left[\ 
{1\over 6} {\rm Tr}\left[\overline{B}S^\mu B\right] 
{\rm Tr}\left[\partial_\mu\pi\right]
\ -\ 
{1\over 6} {\rm Tr}\left[\overline{B}S^\mu B\partial_\mu\pi\right]
\ +\ 
{2\over 3} {\rm Tr}\left[\overline{B}S^\mu \partial_\mu\pi B\right]
\right.
\nonumber\\
& & 
\left.
+\ {1\over 6}\epsilon^{nij} \overline{B}^k_n \ S^\mu \ 
\partial_\mu\chi^{\tilde a \dagger}_i\ _{\tilde a}\Se_{jk}
\ +\ 
{1\over 6}\epsilon_{nij}\  ^{\tilde a}\overline{\Se}^{jk}
\ S^\mu \ 
\partial_\mu\chi_{\tilde a}^i\ B_k^n
\right.
\nonumber\\
& & 
\left.
-\ \sqrt{3\over 8}  
{\rm Tr}\left[\ \overline{B}\ S^\mu \ \partial_\mu\chi^\dagger \ 
\Tr\ \right]
 -\ \sqrt{3\over 8}  
{\rm Tr}\left[\ \overline{\Tr}\ S^\mu\ \partial_\mu\chi \ B\ \right]
\right]
\ +\ ...
\nonumber\\
\left( \overline{\cal B} S^\mu A_\mu {\cal B}\right) & = & 
{1\over f}
\left[\ 
{2\over 3} {\rm Tr}\left[\overline{B}S^\mu B\right] 
{\rm Tr}\left[\partial_\mu\pi\right]
\ -\ 
{2\over 3} {\rm Tr}\left[\overline{B}S^\mu B\partial_\mu\pi\right]
\ -\ 
{1\over 3} {\rm Tr}\left[\overline{B}S^\mu \partial_\mu\pi B\right]
\right.
\nonumber\\
& & 
\left.
+\ {2\over 3}\epsilon^{nij} \overline{B}^k_n \ S^\mu \ 
\partial_\mu\chi^{\tilde a \dagger}_i\ _{\tilde a}\Se_{jk}
\ +\ 
{2\over 3}\epsilon_{nij}\  ^{\tilde a}\overline{\Se}^{jk}
\ S^\mu \ 
\partial_\mu\chi_{\tilde a}^i\ B_k^n
\right]
\ +\ ...
\nonumber\\
\left( \overline{\cal B} S^\mu {\cal B} \right) {\rm str}\left( A_\mu\right)
& = & 
{1\over f}
\left[ {\rm Tr}\left[\  \overline{B}\ S^\mu\  B\ \right]
\ +\ 
{\rm Tr}\left[\  \overline{\Tr}\ S^\mu\  \Tr\ \right]
\ +\ 
{\rm Tr}\left[\  \overline{\Se}\ S^\mu\  \Se\ \right]\ 
\right]\  
{\rm Tr}\left[ \partial_\mu\pi-\partial\tilde\pi \right]
\ +\ ...
\nonumber\\
\left(\overline{\cal B} A^\mu {\cal T}_\mu\right)
& = & 
{1\over f}\sqrt{2\over 3} \epsilon^{xij}
\left[\ 
\overline{B}^k_x\  \partial^\mu\pi^{l}_i\  T_{ljk}
\ +\ 
{1\over\sqrt{3}}
\overline{B}^k_x\  \partial^\mu\chi^{\tilde a \dagger}_i
\ _{\tilde a}\SD_{jk}\ 
\right]
\ +\ ...
\ \ \ ,
\label{eq:intsexp}
\end{eqnarray}
where all the indices that appear in eq.~(\ref{eq:intsexp})
are restricted to be $1,2,3$.
Terms that do not contribute at one-loop level to the magnetic moments
of the octet baryons have not been shown in eq.~(\ref{eq:intsexp}).

The appropriate one-loop diagrams resulting from the Lagrange density
in eqs.~(\ref{eq:lagpi}), (\ref{eq:free}) and (\ref{eq:ints}) 
reproduce both 
the ${\cal O}(m_q^{3/2})$, the ${\cal O}(\alpha_\Phi  m_q^{3/2})$ 
and the ${\cal O}(M_0^2 m_q^{1/2})$ contributions
to the masses of the octet baryons
(the last two coming from hairpin diagrams)
as calculated in LS~\cite{LS96} using quark-line diagrams.
We have not computed the higher order contributions 
to the octet baryon masses, nor the contributions to the decuplet masses,
but expect that we would recover the results of LS~\cite{LS96}.

\section{Magnetic Moments of the Octet Baryons}

The magnetic moments of the octet baryons have been studied extensively in QCD.
Coleman and Glashow (CG)~\cite{CG61}
first considered the moments in the limit of exact $SU(3)$ flavor symmetry,
arriving at six relations between the moments and the $\Sigma^0-\Lambda$
transition moment, known as the Coleman-Glashow relations.
The leading dependence upon the light quark masses, of the form $\sqrt{m_q}$,
was first computed by Caldi and Pagels~\cite{CP74},
and more recently higher order contributions have been computed including the 
decuplet as a dynamical degree of freedom~\cite{JLMS92,MS97,DH98,PRM00}.
In QQCD one expects that
$\sim\sqrt{m_q}$ terms are suppressed as only quarks in the asymptotic
states can participate in meson loops.
However, hairpin interactions, 
and in particular those arising from the operator with coefficient 
$M_0^2$ in eq.~(\ref{eq:lagpi}),
give rise to contributions of the form $\sim M_0^2\log m_q$,
due to the more singular behavior of QQCD in the chiral limit
than QCD.

\subsection{Tree-Level}

The leading order contribution to the magnetic moments of the octet baryons
arises from two dimension-5 operators,
\begin{eqnarray}
{\cal L} & = & 
{e\over 4 M_N} F_{\mu\nu}\ 
\left[\ 
\mu_\alpha\ \left(\ \overline{\cal B}\ \sigma^{\mu\nu}\  {\cal B}\  
{\cal Q}_{\xi+}\ \right)
\ +\ 
\mu_\beta\ \left(\ \overline{\cal B}\ \sigma^{\mu\nu} \ {\cal Q}_{\xi+}\ 
{\cal B}\ \right)
\ \right]
\nonumber\\
{\cal Q}_{\xi+} & = & {1\over 2} 
\left(\ \xi^\dagger {\cal Q}\xi + \xi {\cal Q}\xi^\dagger
\right)
\ \ \ ,
\label{eq:dimfive}
\end{eqnarray}
where
${\cal Q}$ is the electric charge matrix,
${\cal Q}~=~{1\over 3}~{\rm diag}(+2,-1,-1,+2,-1,-1)$.
By considering only the ground floor baryons one can make the identification
with the flavor structure of operators used in QCD
\begin{eqnarray}
{\cal L} & = & 
{e\over 4 M_N} F_{\mu\nu}\ 
\left(\ 
\mu_D\ {\rm Tr}\left[\ \overline{B}\sigma^{\mu\nu} \{Q,B\}\ \right]
\ +\ 
\mu_F\ {\rm Tr}\left[\ \overline{B}\sigma^{\mu\nu} \left[Q,B\right]\ \right]
\ 
\right)
\ +\ ...
\ \ \ ,
\label{eq:magQCD}
\end{eqnarray}
where the ellipses denotes terms involving the meson field,
and find that
\begin{eqnarray} 
\mu_\alpha & = & {2\over 3}\mu_D+2\mu_F
\ \ ,\ \ 
\mu_\beta \ = \ -{5\over 3}\mu_D + \mu_F
\ \ \ .
\end{eqnarray}
Again, one should keep in mind that the values of $\mu_D$ and $\mu_F$ in
QQCD will differ from those in QCD.
The Lagrange density in eq.~(\ref{eq:dimfive}) not only generates magnetic
moments for the octet baryons but also for the entire {\bf 70} dimensional
representation.
\begin{table}[!ht]
\begin{tabular}[h]{cccc} 
 & Baryon & $\mu_i^{\rm tree}$\ (NM) & \\ \hline
& $p$\ ,\ $\Sigma^+$ & ${1\over 3}\mu_D + \mu_F$ & \\
&$n$\ ,\ $\Xi^0$ & $ -{2\over 3}\mu_D$&\\
& $\Sigma^0$ & ${1\over 3}\mu_D$&\\
&$\Sigma^-$\ ,\ $\Xi^-$ & ${1\over 3}\mu_D - \mu_F$&\\
&$\Lambda$ & $-{1\over 3}\mu_D$&\\
&$\Sigma^0-\Lambda$-{\rm transition} & ${1\over\sqrt{3}}\mu_D$ &
\end{tabular}
\caption{
The tree-level contribution to the magnetic moments of the octet baryons 
and the $\Sigma^0-\Lambda$ transition moment
(i.e. the ground floor of the {\bf 70})
in units of nuclear magnetons (NM).
}
\label{table1}
\end{table}
In Table~\ref{table1} we show the tree-level 
magnetic moments of the octet baryons and $\Sigma^0-\Lambda$ 
transition moment in units of nuclear magnetons.
We have not shown the tree-level contributions to the first or higher
floors of the {\bf 70} as they do not contribute to the magnetic 
moments of the octet baryons at one-loop.

As there are only two dimension-5
invariants in QQCD, the same number as in QCD, 
the CG relations~\cite{CG61} between the magnetic moments 
of the octet baryons in the limit of exact $SU(3)_V$ flavor symmetry
persist:
\begin{eqnarray}
\mu_{\Sigma^+} & = & \mu_p
\ \ \  ,\ \ \ 
\mu_{\Sigma^-} + \mu_n \ =\ -\mu_p
\nonumber\\
2\mu_\Lambda & = & \mu_n
\ \ \ ,\ \  \ 
\mu_{\Xi^-} \ =\ \mu_{\Sigma^-}
\nonumber\\
\mu_{\Xi^0} & = & \mu_n
\ \ \  ,\ \  \ 
2\mu_{\Sigma^0\Lambda} \  =\ -\sqrt{3}\mu_n
\ \ \ .
\label{eq:CGtree}
\end{eqnarray}

\subsection{Contributions of the form $\sim\sqrt{m_q}$}

\begin{figure}[!ht]
\centerline{{\epsfxsize=4.0in \epsfbox{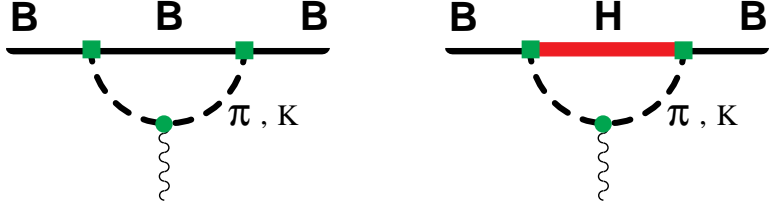}}} 
\vskip 0.15in
\noindent
\caption{\it 
One-loop graphs that give contributions of the form 
$\sim\sqrt{m_q}$
to the magnetic moments of the octet baryons in QCD.
A solid, thick-solid and dashed line denote an 
octet baryon, decuplet baryon, and a meson, respectively.
The wiggly line denotes a photon.
The solid-squares, solid-circles
denote an axial, minimal electromagnetic coupling, respectively.
In QQCD, in addition to the intermediate states involving an octet or
decuplet baryon and a charged pseudo-scalar meson, there are contributions
from ghost-baryons, the $\Tr$'s, $\Se$'s and $\SD$'s, 
and fermionic pseudo-scalar mesons, 
the $\chi_{\pi^\pm}$ and $\chi_{K^\pm}$.
}
\label{fig:sqrtmq}
\vskip .2in
\end{figure}

In QCD the leading contributions to the magnetic moments that depend upon the 
light-quark masses arise from the one-loop graphs shown in
fig.~\ref{fig:sqrtmq}.
The photon couples to a charged meson which is coupled to the baryon via
axial interactions.  Graphs with octet baryons in the intermediate state
give contributions of the form $\sim\sqrt{m_q}$, whereas those with decuplet
intermediate states have a more complicated functional dependence upon 
$m_q$ due to the presence of the decuplet-octet mass splitting, $\Delta$.
In QQCD the diagrams shown in fig.~\ref{fig:sqrtmq}
also generate contributions of the form
$\sim\sqrt{m_q}$, but in addition to intermediate states of
the ground floor of baryons
and pseudo-scalar mesons, $\pi^\pm$ and $K^\pm$, there are
contributions from intermediate states of
ghost-baryons, the $\Tr$'s, $\Se$'s and $\SD$'s, 
and fermionic mesons, the $\chi_{\pi^\pm}$ and $\chi_{K^\pm}$.

Following Ref.~\cite{JLMS92}, we write the contribution to the magnetic moments
from these diagrams as
\begin{eqnarray}
\delta\mu_i & = & {M_N\over 4\pi f^2}\sum_{X=\pi ,K}\  
\left[\ 
\beta^{(X)}_i\ m_X
\ +\ 
\beta^{\prime (X)}_i {\cal F}(m_X,\Delta,\mu)
\ \right]
\ \ \ ,
\label{eq:sqrtmqform}
\end{eqnarray}
which is true in both QCD and QQCD
(Ref.~\cite{JLMS92} uses $f\sim 94~{\rm MeV}$ and hence the coefficients
in eq.~(\ref{eq:sqrtmqform}) differ from those of Ref.~\cite{JLMS92}
by  a factor of $2$).
The $\beta^{(X)}_i$ contributions arise from 
diagrams with baryons in the {\bf 70} dimensional representation
in the intermediate state, while
the $\beta^{\prime (X)}_i$ contributions arise from 
diagrams with baryons in the {\bf 38} dimensional representation
in the intermediate state.
The loop-function ${\cal F}(m_X,\Delta,\mu)$ is defined to be~\cite{JLMS92} 
\begin{eqnarray}
\pi {\cal F}(m,\Delta,\mu)
& = & \sqrt{\Delta^2-m^2}\log\left({\Delta-\sqrt{\Delta^2-m^2+i\epsilon}
\over \Delta+\sqrt{\Delta^2-m^2+i\epsilon}}\right)
\ -\ \Delta\log\left({m^2\over\mu^2}\right)
\ \ \ ,
\end{eqnarray}
where $\mu$ is the renormalization scale, and
${\cal F}(m,0,\mu)=m$.

\begin{table}[!ht]
\begin{tabular}[h]{ccccc}
& \multicolumn{2}{c}{ QCD }  
& \multicolumn{2}{c}{ QQCD } \\ 
Baryon  & $\beta^{(\pi)}$ & $\beta^{(K)}$ & $\beta^{(\pi)}$ & $\beta^{(K)}$
\\ \hline
$p$ & $-(D+F)^2$ & $-{2\over 3} D^2 - 2 F^2$ &$ -{4\over 3}D^2$ & $0$\\
$n$ & $(D+F)^2$ & $-(D-F)^2$ &  $ {4\over 3}D^2$ & $0$\\
$\Sigma^+$ & $-{2\over 3} D^2 - 2 F^2$ & $-(D+F)^2$ & $0$ & $ -{4\over 3}D^2$\\
$\Sigma^0$ & $0$ & $-2 D F$ & $0$ & $-{2\over 3} D^2$\\
$\Sigma^-$ & ${2\over 3} D^2 + 2 F^2$ & $(D-F)^2$ & $0$ & $0$\\
$\Lambda$ & $0$ & $2 D F $ & $0$ & ${2\over 3} D^2 $\\
$\Xi^-$ & $(D-F)^2$ &${2\over 3} D^2 + 2 F^2$ & $0$ & $0$\\
$\Xi^0$ & $-(D-F)^2$ & $(D+F)^2$ &  $0$ &$ {4\over 3}D^2$\\
$\Sigma^0\Lambda$ & $-{4\over\sqrt{3}} D F $ & $-{2\over\sqrt{3}} D F $ &
$-{4\over 3\sqrt{3}} D^2$ & $-{2\over 3\sqrt{3}} D^2$
\end{tabular}
\caption{
The coefficients $\beta^{(\pi)}$ and $\beta^{(K)}$ 
in QCD and QQCD arising from 
the one-loop graphs shown in fig.~\ref{fig:sqrtmq}.
}
\label{table2}
\end{table}
\begin{table}[!ht]
\begin{tabular}[h]{ccccc}
& \multicolumn{2}{c}{ QCD }  
& \multicolumn{2}{c}{ QQCD } \\ 
Baryon  & $\beta^{\prime (\pi)}$ & $\beta^{\prime(K)}$ 
& $\beta^{\prime (\pi)}$ & $\beta^{\prime (K)}$
\\ \hline
$p$ & $-{2\over 9}{\cal C}^2$ & ${1\over 18}{\cal C}^2 $ &$ -{1\over 6}{\cal
  C}^2 $ & $0$\\
$n$ & ${2\over 9}{\cal C}^2$ & ${1\over 9}{\cal C}^2 $ &$ {1\over 6}{\cal
  C}^2 $ & $0$\\
$\Sigma^+$ & ${1\over 18}{\cal C}^2$ & $-{2\over 9}{\cal C}^2 $ & $0$ &
$-{1\over 6}{\cal  C}^2 $\\ 
$\Sigma^0$ & $0$ & $-{1\over 6}{\cal C}^2 $ & $0$ &
$-{1\over 12}{\cal  C}^2 $\\ 
$\Sigma^-$ &$-{1\over 18}{\cal C}^2$ & $-{1\over 9}{\cal C}^2 $ & $0$ &
$0$\\ 
$\Lambda$ & $0$ & ${1\over 6}{\cal C}^2 $ & $0$ &
${1\over 12}{\cal  C}^2 $\\ 
$\Xi^-$ & $-{1\over 9}{\cal C}^2$ & $-{1\over 18}{\cal C}^2 $ & $0$ &
$0$\\ 
$\Xi^0$ &  ${1\over 9}{\cal C}^2$ & ${2\over 9}{\cal C}^2 $ & $0$ &
${1\over 6} {\cal C}^2$\\ 
$\Sigma^0\Lambda$ & $-{1\over 3\sqrt{3}}{\cal C}^2$ &  
$-{1\over 6\sqrt{3}}{\cal C}^2$ & $-{1\over 6\sqrt{3}}{\cal C}^2$ 
& $-{1\over 12\sqrt{3}}{\cal C}^2$
\end{tabular}
\caption{
The coefficients $\beta^{\prime (\pi)}$ and $\beta^{\prime (K)}$ 
in QCD and QQCD arising from 
the one-loop graphs shown in fig.~\ref{fig:sqrtmq}.
}
\label{table3}
\end{table}
The computed values of the coefficients $\beta^{(\pi)}$, $\beta^{(K)}$, 
$\beta^{\prime (\pi)}$ and  $\beta^{\prime (K)}$ 
that appear in eq.~(\ref{eq:sqrtmqform})
are shown in Table~\ref{table2} and Table~\ref{table3}
for both QCD and QQCD.
It is interesting to note that, unlike QCD, there is no contribution from the 
F-type axial interaction in QQCD.
In QCD, there are three relations between the moments of the octet 
baryons that 
hold in the presence of the $\sim\sqrt{m_q}$ contributions, 
as first pointed out by Caldi and Pagels (CP)~\cite{CP74} for octet baryons 
in the intermediate state, and subsequently shown to be valid 
for decuplet baryons in the intermediate state~\cite{JLMS92},
\begin{eqnarray}
\mu_{\Sigma^+} & = & -2\mu_\Lambda - \mu_{\Sigma^-}
\ \ ,\ \ 
\mu_\Lambda-\sqrt{3}\mu_{\Sigma^0\Lambda} \ = \ \mu_{\Xi^0}+\mu_n
\ \ ,\ \ 
\mu_{\Xi^0} + \mu_{\Xi^-} +\mu_n = 2\mu_\Lambda -\mu_p
\ \ \ ,
\end{eqnarray}
which are found to hold at the $\sim 10\%$-level in nature~\cite{JLMS92}.
This is to be compared with the CG-relations~\cite{CG61}, 
that are found to hold at the $\sim 30\%$-level, consistent with 
naive expectations.
The CP-relations are also true in QQCD
when only the $\sim\sqrt{m_q}$ contributions are considered, 
as can be seen with ease in
Table~\ref{table2} and Table~\ref{table3}.
It is not surprising to see that the size of the loop corrections in QQCD are
smaller than their counterparts in QCD.
After all, quenching removes quark loop
contributions. A consequence of this is that the loop expansion for the 
magnetic moments is expected to be more convergent, and hopefully valid for
higher quark masses in QQCD than in QCD, as it should be for all observables.

\subsection{Contributions of the form $M_0^2 \log m_q$ : Hairpins}

\begin{figure}[!ht]
\centerline{{\epsfxsize=3.5in \epsfbox{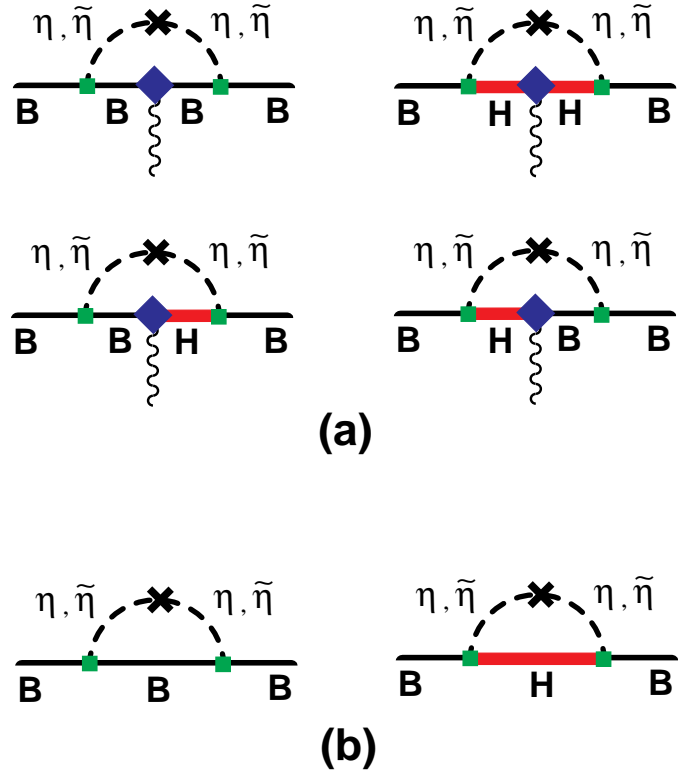}}} 
\vskip 0.15in
\noindent
\caption{\it 
The one-loop graphs that give contributions of the form 
$\sim M_0^2 \log m_q$
to the magnetic moments of the octet baryons in QQCD.
A solid, thick-solid, dashed line
denotes an octet baryon, decuplet baryon, meson, respectively. 
A wiggly line denotes a photon.
A cross on a meson line corresponds to an insertion 
of the hairpin interaction
with coefficient $M_0$,
while the solid-diamonds, solid-squares 
correspond to magnetic moment, axial interactions, 
respectively.
Diagrams (a) are vertex corrections while diagrams 
(b) give rise to  wavefunction renormalization.
}
\label{fig:HP}
\vskip .2in
\end{figure}

Unlike QCD, in QQCD there are contributions to the magnetic moments of the 
baryons of the form $\sim M_0^2 \log m_q$
from the diagrams in fig.~\ref{fig:HP}, where $M_0$ is one of the  
hairpin interactions defined in eq.~(\ref{eq:lagpi}).
Of course, there are other hairpin interactions that will contribute, 
but this one gives the terms that are most singular in the chiral limit.
For these hairpin contributions, only the ground floor baryons contribute,
but both $\eta_q$ and $\tilde\eta_q$ mesons can appear in the loop.
As discussed in Ref.~\cite{Chow}, the contributions from the axial hairpin
interaction, with coefficient $\gamma$, in such diagrams vanish, 
leaving only 
contributions from the $F$, $D$ and ${\cal C}$ interactions.

The wavefunction renormalization factors  arising from the diagrams shown in
fig.~\ref{fig:HP}(b) are 
\begin{eqnarray}
Z_\psi^{N} & = & 
1 - {m_0^2\over 16\pi^2 f^2}\ (D-3F)^2 \ \overline{I}_{uu}
\nonumber\\
Z_\psi^\Sigma & = & 
1 - {m_0^2\over 16\pi^2 f^2}
\left[ 
4 F^2 \overline{I}_{uu} + (D-F)^2  \overline{I}_{ss}
+ 4 F (F-D) \overline{I}_{us}
+ {2 {\cal C}^2\over 9} \left( \overline{I}_{ss}^{\Delta\Delta}
+\overline{I}_{uu}^{\Delta\Delta} - 2 \overline{I}_{us}^{\Delta\Delta}
\right)
\right]
\nonumber\\
Z_\psi^\Xi & = & 
1 - {m_0^2\over 16\pi^2 f^2}
\left[ 
4 F^2 \overline{I}_{ss} + (D-F)^2  \overline{I}_{uu}
+ 4 F (F-D) \overline{I}_{us}
+ {2 {\cal C}^2\over 9} \left( \overline{I}_{ss}^{\Delta\Delta}
+\overline{I}_{uu}^{\Delta\Delta} - 2 \overline{I}_{us}^{\Delta\Delta}
\right)
\right]
\nonumber\\
Z_\psi^\Lambda & = & 
1 - {m_0^2\over 16\pi^2 f^2}
\left[ 
{4\over 9}(2D-3F)^2  \overline{I}_{uu}
-{4\over 9}(2D^2 + 3 DF-9F^2) \overline{I}_{us}
+ {1\over 9}(D+3F)^2 \overline{I}_{ss}
\right]
\  ,
\label{eq:zpsi}
\end{eqnarray}
where the contribution to an S-matrix element is
$\sim \sqrt{Z_\psi^{i}}\ \sqrt{Z_\psi^{f}}\  \Gamma^{\rm tree}$, where
$\Gamma^{\rm tree}$ is the tree-level vertex and $i,f$ denote the initial and
final state baryon, respectively.
The integrals that appear in eq.~(\ref{eq:zpsi}) are defined to be 
\begin{eqnarray}
\overline{I}_{q q^\prime} & = & 
\overline{I}(m_{\eta_q}, m_{\eta_{q^\prime}}, 0, 0, \mu)
\ ,\ 
\overline{I}_{q q^\prime}^\Delta =
\overline{I}(m_{\eta_q}, m_{\eta_{q^\prime}}, \Delta, 0, \mu)
\ ,\ 
\overline{I}_{q q^\prime}^{\Delta\Delta} =  
\overline{I}(m_{\eta_q}, m_{\eta_{q^\prime}}, \Delta, \Delta, \mu)
\ ,
\end{eqnarray}
where
\begin{eqnarray}
& & \overline{I}(m_1, m_2, \Delta_1, \Delta_2, \mu) 
=  
{
\left[ \ Y(m_1,\Delta_1, \mu)+ Y(m_2,\Delta_2, \mu)
- Y(m_1,\Delta_2, \mu)-Y(m_2,\Delta_1, \mu)
\ \right]
\over
[\Delta_1-\Delta_2][m_1^2-m_2^2]}
\nonumber\\
& & \qquad = \ 
-i {16\pi^2\over 3}
\int {d^n q\over (2\pi)^n}
{q^2-(v\cdot q)^2\over
[v\cdot q-\Delta_1 + i \epsilon][v\cdot q-\Delta_2 + i \epsilon]
[q^2-m_1^2 + i \epsilon][q^2-m_2^2 + i \epsilon]}
\ \ \ ,
\label{eq:Ibardef}
\end{eqnarray}
with
\begin{eqnarray}
Y(m,\Delta, \mu) & = & 
\left[m^2-{2\over 3}\Delta^2\right]\Delta\log \left({m^2\over\mu^2}\right)
\ +\ 
{2\over 3}\left[\Delta^2-m^2\right]^{3\over 2}
\log\left({\Delta-\sqrt{\Delta^2-m^2+i\epsilon}\over
\Delta+\sqrt{\Delta^2-m^2+i\epsilon}}\right)
\ .
\end{eqnarray}
The integral in eq.~(\ref{eq:Ibardef}) has nice limits, e.g.
\begin{eqnarray}
\overline{I}(m_1,m_2,0,0, \mu) & = & 
{m_1^2\log \left({m_1^2\over\mu^2}\right)
-m_2^2\log \left({m_2^2\over\mu^2}\right)\over m_1^2-m_2^2}
\ +\ ...
\nonumber\\
\overline{I}(m,m,0,0, \mu) & = &   \log\left( {m^2\over\mu^2}\right)
\ +\ ...
\ ,
\label{eq:lims}
\end{eqnarray}
where the ellipses denote terms that are analytic in $m_q$.
In computing the loop graphs we have performed the spin algebra in four
dimensions.

In the diagrams shown in fig.~\ref{fig:HP} one sees that there are
contributions from the magnetic moments of the decuplet baryons, 
and from the transition magnetic moment between the decuplet and octet baryons.
The Lagrange density that describes these interactions is~\cite{JLMS92}
\begin{eqnarray}
{\cal L} & = & 
-i\sqrt{3\over 2} \mu_T {e\over 2 M_N} F_{\mu\nu} 
\left[\left(\overline{{\cal B}}S^\mu {\cal Q}_{\xi+} 
{\cal T}^\nu\right)\ +\ {\rm
    h.c.}\right]
\ -\ 
i\  3\  \mu_C {e\over 2 M_N} F_{\mu\nu} 
\left(\overline{{\cal T}}^\mu {\cal Q}_{\xi+}  {\cal T}^\nu\right)
\nonumber\\
& = & i\mu_T\ {e\over 2 M_N}F_{\mu\nu}
\left( \epsilon_{ijk} 
Q^i_l \overline{B}^j_m S^\mu T^{\nu, klm}
 \ +\ 
\epsilon^{ijk} Q^l_i 
\overline{T}^\mu_{klm} S^\nu B^m_j\right)
\ -\ i \mu_C\ {e\over M_N} q_i \overline{T}^\mu_{i} T^\nu_i F_{\mu\nu}
\nonumber\\
& & \ +\ ...
\ \ \ ,
\end{eqnarray}
where $q_i$ is the charge of the $i^{th}$ decuplet baryon, and the ellipses
denote terms involving meson fields.
The expressions that one obtains for the hairpin contribution to the magnetic
moment of each baryon can be written as
\begin{eqnarray}
\delta\mu_i & = & {M_0^2\over 16\pi^2 f^2}
\left[
\ -{4\over 3}\ \mu_i^{\rm tree}\ w^{(8,8)}_i\ +\ 
{\cal C}\ \mu_T\ 
w^{(8,10)}_i
\ +\ {\cal C}^2\left(
\overline{I}_{ss}^{\Delta\Delta}
+\overline{I}_{uu}^{\Delta\Delta}
- 2 \overline{I}_{us}^{\Delta\Delta}\right)\ w^{(10,10)}_i
\ \right]
\ ,
\label{eq:HPdef}
\end{eqnarray}
where $w^{(8,8)}_i$ is the contribution from intermediate states with octet
baryons,
 $w^{(10,10)}_i$ is the contribution from intermediate states with decuplet
baryons,
and 
 $w^{(8,10)}_i$ is the contribution from intermediate states with 
both octet and decuplet baryons.
\begin{table}[!ht]
\begin{tabular}[h]{cc} 
Baryon & $w^{(8,8)}_i$ \\ \hline
$p\ ,\ n$ & $(D-3 F)^2\  \overline{I}_{uu}$\\
$\Sigma^+\ ,\ \Sigma^0\ ,\ \Sigma^-$ & 
$4F^2\overline{I}_{uu}+ (D-F)^2 \overline{I}_{ss}
+4F(F-D)\overline{I}_{us}$ \\
$\Lambda$ & 
${1\over 9}\ \left(
4(2D-3F)^2 \overline{I}_{uu} - 4(2D^2+3DF-9F^2) \overline{I}_{us}
+(D+3F)^2 \overline{I}_{ss}
\right)$ \\
$\Xi^0\ ,\ \Xi^-$ & 
$4 F^2 \overline{I}_{ss} + (D-F)^2  \overline{I}_{uu} + 4 F(F-D)
\overline{I}_{us}$ \\
$\Sigma^0\rightarrow\Lambda$ & 
$\left(4F^2 -{8\over 3}DF + {2\over 3}D^2\right)\overline{I}_{uu}
+ 
\left(F^2 -{2\over 3}DF + {1\over 3}D^2\right)\overline{I}_{ss}
+4 F \left(F-{2\over 3}D\right) \overline{I}_{us}$
\end{tabular}
\caption{
Hairpin contributions from graphs involving octet baryons in the intermediate
state.  The coefficient $w^{(8,8)}_i$ is defined in eq.~(\ref{eq:HPdef}).
}
\label{table4}
\end{table}
\begin{table}[!ht]
\begin{tabular}[h]{ccc} 
Baryon & $w^{(8,10)}_i$  & $w^{(10,10)}_i$\\ \hline
$p\ ,\ n$ & $0$ & $0$\\
$\Sigma^+$ 
& ${4\over 27}\left( 
(F-D)\left(\overline{I}_{us}^{\Delta}-\overline{I}_{ss}^{\Delta}\right)
\ + \ 
2 F \left(\overline{I}_{uu}^{\Delta}-\overline{I}_{us}^{\Delta}\right)\right)$
& 
$-{2\over 9}\ \ \mu_{\Sigma^+}^{\rm tree} \ +\ 
{20\over 81}\ \mu_C\ $ \\
$\Sigma^0$ & 
${2\over 27}\left( 
(F-D)\left(\overline{I}_{us}^{\Delta}-\overline{I}_{ss}^{\Delta}\right)
\ + \ 
2 F \left(\overline{I}_{uu}^{\Delta}-\overline{I}_{us}^{\Delta}\right)
\right)$
& $-{2\over 9}\ \ \mu_{\Sigma^0}^{\rm tree} $ \\
$\Sigma^-$ & $0$ &
$-{2\over 9}\ \ \mu_{\Sigma^-}^{\rm tree} \ -\ 
{20\over 81}\ \mu_C$ \\
$\Lambda$ & $0$ &
$0$ \\
$\Xi^0$ 
& ${4\over 27}\left( 
(F-D)\left(\overline{I}_{uu}^{\Delta}-\overline{I}_{us}^{\Delta}\right)
\ + \ 
2 F \left(\overline{I}_{us}^{\Delta}-\overline{I}_{ss}^{\Delta}\right)\right)$
& 
$-{2\over 9}\ \ \mu_{\Xi^0}^{\rm tree}$ \\
$\Xi^-$ & $0$ &
$-{2\over 9}\ \ \mu_{\Xi^-}^{\rm tree} 
 \ -\ {20\over 81}\ \mu_C$ \\
$\Sigma^0\rightarrow\Lambda$ &  $0$ &
$-{1\over 9}\ \ \mu_{\Sigma^0\Lambda}^{\rm tree} $
\end{tabular}
\caption{
Hairpin contributions from graphs involving both  octet 
and decuplet
baryons in the intermediate
state.  The coefficients $w^{(8,10)}_i$ and  $w^{(10,10)}_i$
are defined in eq.~(\ref{eq:HPdef}).
}
\label{table5}
\end{table}
The expressions that we have computed for the $w^{(r,s)}_i$ are shown in 
Table~\ref{table4} and Table~\ref{table5}.

Corrections of the form $\sim\sqrt{m_q}$ and $\sim m_q\log m_q$ have been
computed in QCD and the counterterms required to render the theory finite have 
been written down~\cite{JLMS92}.
At this order there is one relation between the magnetic
moments,
\begin{eqnarray}
6\mu_\Lambda + \mu_{\Sigma^-} - 4\sqrt{3}\mu_{\Sigma^0\Lambda}
\ =\ 
4\mu_n-\mu_{\Sigma^+} + 4\mu_{\Xi^0}
\ \ \ .
\label{eq:JLMS}
\end{eqnarray}
that is found to be satisfied at $\sim 5\%$-level in nature.
However, in QQCD we have been unable to find any $SU(3)$ relation
between the magnetic moments when both the $\sim\sqrt{m_q}$ and 
$\sim M_0^2\log m_q$ contributions are considered.
Of course, the trivial $SU(2)$ relation between the $\Sigma^{\pm,0}$
moments 
$\mu_{\Sigma^+}\ +\ \mu_{\Sigma^-}\ -\ 2 \mu_{\Sigma^0}=0$
still holds.

\subsection{The Nucleon}

It is worth considering the chiral expansion of the 
nucleon magnetic moment explicitly.
We have found that the magnetic moment of the proton is 
\begin{eqnarray}
\mu_p & = & 
\left[{\mu_D\over 3}+\mu_F\right]
\left[1 - {M_0^2(D-3 F)^2\over 12\pi^2 f^2} 
\log \left({m_\pi^2\over\mu^2}\right)\right]
-
{M_N\over 24\pi f^2}\left[ 
8 D^2 m_\pi +{\cal C}^2 {\cal F}(m_\pi,\Delta,\mu)
\right]
\  ,
\label{eq:proton}
\end{eqnarray}
where we have shown only the terms non-analytic in  $m_q$.
It has all the features we expect to find in QQCD.
Kaon loops are absent as there are no strange quarks in the 
initial or final states.
There are still pion loop contributions as the up and down quarks in the 
initial and final states can be routed through a meson-loop
diagram without requiring the presence of a sea-quark.
Further, if we insert the QCD values of the axial 
couplings~\cite{chiralN}, $F$ and $D$, we find that the pion loop contribution
is somewhat smaller in QQCD than QCD. It will be interesting to learn the 
values of $F$ and $D$ in QQCD to see if this feature is actually present.

\begin{figure}[!ht]
\centerline{{\epsfxsize=3.5in \epsfbox{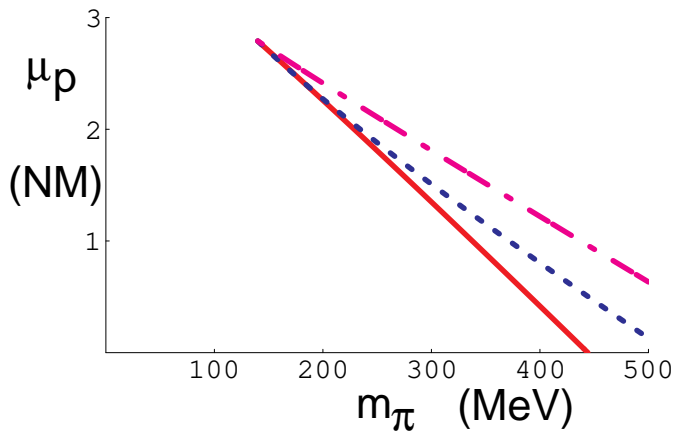}}}
\vskip 0.15in
\noindent
\caption{\it 
The magnetic moment of the proton verses the pion mass, 
with the kaon mass fixed to its physical value.
The solid curve corresponds to QCD with only the 
$\sim\sqrt{m_q}$ light-quark mass dependence included.
The dashed curve corresponds to QQCD with
$M_0=750~{\rm MeV}$~\protect\cite{KFMOU94}.
To demonstrate the sensitivity of the magnetic moment to 
$M_0$, the dot-dashed 
curve corresponds to $M_0=400~{\rm MeV}$~\protect\cite{BDEIT01}.
}
\label{fig:proton}
\vskip .2in
\end{figure}

In order to gain some sort of understanding of how the QQCD extrapolation 
compares with the QCD extrapolation, we use the QCD values for the constants,
that appear in eq.~(\ref{eq:proton}),
$F=0.5$, $D=0.8$~\cite{FMJM98}, $|{\cal C}|=1.8$~\cite{BSS92}, 
$f=132~{\rm MeV}$, and the physical value
of the $\Delta$-$N$ mass splitting.
Choosing to renormalize at the scale $\mu\sim 1~{\rm GeV}$, and
keeping the kaon mass fixed at its physical value we vary the pion mass
and determine the proton magnetic moment.
The tree-level magnetic moment is fixed so that 
at one-loop it takes its experimental value for the 
physical pion mass, 
while $M_0\sim 750~{\rm MeV}$ 
is taken from the work in Ref.~\cite{KFMOU94} 
(discussed in Ref.~\cite{G94}).
The magnetic moment of the proton verses the pion mass is shown in 
fig.~\ref{fig:proton}.
One sees that the extrapolation in QQCD is not so different to that in QCD.
The reduction in the $\sim\sqrt{m_q}$ terms is somewhat compensated by the
contribution from the hairpin diagrams.
If one uses the values of the axial couplings extracted at one-loop order,
$F\sim 0.4$, $D\sim 0.6$, $|{\cal C}|\sim 1.4$~\cite{BSS92,SW96}, 
the slope of each curve is reduced.
It is clear from fig.~\ref{fig:proton} that 
the nucleon magnetic moment depends strongly upon $m_q$ 
at one-loop order
(for these particular values of couplings), 
and that a higher order calculation in QQCD is likely  required, 
as was the case in QCD~\cite{MS97}.
Further, with these parameters, it would appear that an 
extrapolation from a pion mass of $\sim 250~{\rm MeV}$ or higher 
is unreliable without a higher order calculation.

\section{Conclusions}

We have computed the contributions of the form 
$\sim M_0^2\log m_q$ and $\sim \sqrt{m_q}$
to the magnetic moments of the octet baryons
in quenched chiral perturbation theory.
The sickness of QQCD reveals itself yet again in the chiral expansion, 
with hairpin interactions providing the formally dominant correction
near the chiral limit, giving a completely different chiral
behavior than QCD.
In performing this calculation we found it convenient to introduce 
ghost-baryon 
fields to describe the first, second and third floors of the 
{\bf 70} and {\bf 38} dimensional representations of $SU(3|3)$, 
the implementation of which  we have detailed in the text.

In order to use our results, not only must the magnetic moments of the baryons
be determined from the lattice, but also the quenched 
axial coupling constants~\footnote{The 
quenched chiral corrections to these quantities have been previously 
computed~\cite{Kim2}.}, $F$, $D$ and ${\cal C}$,
and the meson decay constant, $f$.
Such calculations 
will enable an extrapolation of quenched lattice data from the lattice
masses to the physical masses to make a prediction 
for the baryon magnetic moments in QQCD.

\bigskip\bigskip

\acknowledgements

I would like to thank Steve Sharpe for many useful discussions.
Also, I would like to thank Aneesh Manohar and Jiunn-Wei Chen 
for useful comments on the manuscript.
This work is supported in
part by the U.S. Dept. of Energy under Grants No.  DE-FG03-97ER4014.

\end{document}